\documentclass[prl,aps,floats,twocolumn]{revtex4}

\usepackage[dvips]{graphicx}
\usepackage{amssymb}
\usepackage{amsmath}

\usepackage{color}

\begin{document}

\title{Perfect porcupines: ideal networks for low frequency
  gravitational wave astronomy}

\author{Latham Boyle}

\affiliation{Perimeter Institute for Theoretical Physics, Waterloo,
  Ontario, Canada \\ Canadian Institute for Theoretical Astrophysics,
  Toronto, Ontario, Canada }

\date{March 2010}
                            
\begin{abstract}
  Perfect porcupines are specially-configured networks of
  gravitational wave detectors, in the limit that the individual
  detectors and the distances between them are short relative to the
  gravitational wavelengths of interest.  They have beautiful
  properties which make them ideal gravitational wave telescopes.  I
  present the most important cases explicitly.  For a network of
  one-arm detectors (like ``AGIS'' \cite{Dimopoulos:2008sv}), the
  minimal perfect porcupine has 6 detectors, oriented along the 6
  diameters of a regular icosahedron.  For a network of two-arm
  detectors (like the equal-arm Michelson interferometers LIGO/VIRGO)
  the minimal perfect porcupine is a certain 5 detector configuration.
\end{abstract}
\maketitle 

\section{Introduction}

In this decade, we hope and expect that gravitational waves will be
directly detected for the first time, and a new field will be born:
gravitational wave astronomy.  Attention will shift, from individual
gravitational wave detectors, to neworks of multiple detectors which
function together as a gravitational wave {\it telescope}.  (For an
introduction to gravitational wave detector networks, see
\cite{Wen:2010cr} and references therein.)  Here I consider the regime
in which (by choice or necessity) the individual detectors and the
distances between them are short relative to the gravitational
wavelengths of interest, so that the arms of the various detectors in
the network may be thought of as emanating from nearly the same point
in space, like the fanned quills of a frightened porcupine.  In this
limit, the relative {\it positions} of the detectors are unimportant,
and it is their relative {\it orientations} that matter.  For special
configurations (particular choices for the number of detectors and
their orientations), such networks exhibit beautiful properties that
make them, in certain respects, ideal astronomical instruments.  These
``perfect porcupines'' are the subject of this Letter.

Motivation comes from recent developments in gravitational wave
detection.  The ground based LIGO/VIRGO detectors \cite{LIGO} have
successfully reached their (``initial'' or ``stage I'') design
sensitivity and will be upgraded over the next several years to a
higher (``advanced'' or ``stage II'') sensitivity which will likely
allow them to detect gravitational waves.  But the initial LIGO/VIRGO
experiments are only sensitive to high frequency ($30~{\rm Hz}\lesssim
f\lesssim 10^{3}~{\rm Hz}$) radiation; they are analogous to gamma ray
telescopes, in that their expected signal consists of occasional brief
bursts which represent only the high frequency tip of the
astrophysical gravitational wave spectrum.  It would be tremendously
valuable to have complementary detectors (analogous to optical or
radio telescopes) to study the rich array of sources and physical
effects at lower frequencies ($f\lesssim10~{\rm Hz}$).  To this end,
several space missions have been proposed; but, unfortunately, there
is uncertainty about when and whether the first generation mission
(LISA \cite{LISA}, optimized near $f\sim10^{-3}~{\rm Hz}$) will
launch, and even greater uncertainty surrounding the proposed second
generation missions (BBO \cite{BBO} and DECIGO \cite{DECIGO},
optimized near $f\sim0.3~{\rm Hz}$).  Thus, it makes sense to give
some consideration to the possibility of pursuing low frequency
($f\lesssim10~{\rm Hz}$) gravitational wave astronomy from the ground
-- as best we can, and hopefully in parallel with the space missions
mentioned above.  But then we quickly find ourselves in the
``porcupine'' regime, because frequencies $f\lesssim50~{\rm Hz}$
correspond to wavelengths longer than the radius of the Earth!

Recent developments have invigorated the possibility of gravitational
wave astronomy at low frequencies ($f\lesssim10~{\rm Hz}$) from the
ground.  In particular, there are ideas for how to design detectors
that are less susceptible to the two noise sources (seismic noise and
gravity gradient noise) that limit LIGO/VIRGO at low frequencies.  One
idea is to operate underground, where seismic noise and gravity
gradient noise are intrinsically weaker and, moreover, may be more
effectively monitored (with seismometers) and removed
\cite{Harms:2009bz}.  A second idea is gravitational wave detection
using atom interferometry \cite{Dimopoulos:2008sv, Hohensee:2010mw,
  Yu:2010ss}, which may be more immune to seismic noise, and has many
possible avenues for future development.  In the Discussion, I suggest
another idea which may be relevant to gravity gradient noise in
certain contexts.

\section{Properties of Perfect Porcupines}

In this section I define perfect porcupines and highlight some of
their good features.  In the next two sections, I present the most
important perfect porcupines explicitly.

First let me fix notation.  My fourier conventions are:
\begin{equation}
   g(t)\!=\!\int_{-\infty}^{+\infty}\!\!\!\!\!\!\!df\,{\rm e}^{-2\pi i ft}\tilde{g}(f),\;\;\;
    \tilde{g}(f)\!=\!\int_{-\infty}^{+\infty}\!\!\!\!\!\!\!dt\,{\rm e}^{+2\pi i ft}g(t).
\end{equation}
I work in transverse-traceless (TT) gauge \cite{MTW}.  The lower case
latin indices $\{i,j\}$ label the 3 spatial directions: $i,j=1,2,3$.
The upper case latin indices $\{A,B\}$ label the 2 gravitational wave
polarizations: $A,B=1,2$.  The lower case greek indices
$\{\alpha,\beta\}$ label the $N$ detectors in the network:
$\alpha,\beta=1,\ldots,N$.  I use the Einstein summation convention:
repeated indices (one upper and one lower) are to be summed.  Hats
denote unit vectors.

When gravitational waves reach us from a distant astronomical source,
they appear as a gravitational plane wave travelling in the $\hat{n}$
direction
\begin{equation}
  h_{ij}^{}(\tau)=\sum_{A=1,2}h_{A}^{}(\tau)P_{ij}^{A}(\hat{n}),
\end{equation}
where the two polarization waveforms $h_{A}(\tau)$ are arbitrary
functions of $\tau=t-\hat{n}\cdot\vec{x}$, and the polarization
tensors $P_{ij}^{A}(\hat{n})$ form an orthonormal basis on the
2-dimensional space of symmetric, transverse, traceless $3\times3$
matrices:
\begin{subequations}
  \begin{eqnarray}
    \label{symmetric}
    P_{ij}^{A}(\hat{n})-P_{ji}^{A}(\hat{n})&=&0 \\
    \label{transverse}
    \hat{n}^{i}P_{ij}^{A}(\hat{n})&=&0 \\
    \label{traceless}
    \delta^{ij}P_{ij}^{A}(\hat{n})&=&0 \\
    \label{orthonormal}
    \big[P^{ij}_{B}(\hat{n})\big]^{\ast}P_{ij}^{A}(\hat{n})&=&\delta^{A}_{B}.
 \end{eqnarray}
\end{subequations}

The output $s_{\alpha}^{}(t)$ of detector $\alpha$ has two parts,
gravitational wave signal $h_{\alpha}^{}(t)$ and noise
$n_{\alpha}^{}(t)$:
\begin{equation}
  s_{\alpha}^{}(t)=h_{\alpha}^{}(t)+n_{\alpha}^{}(t).
\end{equation}

Let us first consider $h_{\alpha}^{}(t)$ and the information it
contains.  In the porcupine limit, all of the detectors in the network
live at essentially the same spatial location
($\vec{x}\approx\vec{0}$), and hence only measure $h_{ij}(t,\vec{x})$
at that point: $h_{ij}(t,\vec{0})$.  The gravitational wave signal is
\begin{equation}
    h_{\alpha}(\tau)=\int_{-\infty}^{+\infty}dT\,
    W_{\alpha}(T)h_{ij}(\tau-T)A_{\alpha}^{ij}
\end{equation}
where the window function $W_{\alpha}(T)$ describes the temporal
response of detector $\alpha$, and the ``antenna pattern''
$A_{\alpha}^{ij}$ [normalized as
$(A_{\alpha\,ij})^{\ast}A_{\alpha}^{ij}=1$] is a projector describing
its spatial orientation.  Since each detector $\alpha$ is sensitive to
a particular linear combination $A_{\alpha}^{ij}h_{ij}$ of the
components of $h_{ij}$, the collection of detectors in the network
measure linearly independent combinations if and only if the
determinant of the $N\times N$ grammian matrix of the antenna
projectors is non-zero:
\begin{equation}
  \label{grammian}
  {\rm Det}[A_{\alpha}^{ij}A_{ij}^{\beta}]\neq0.
\end{equation}
We can have up to $6$ linearly independent antenna projectors, and in
this maximal case the network measures all 6 components of
$h_{ij}(t,\vec{0})$.  Then, even though these measurements are
confined to $\vec{x}=\vec{0}$, they are enough to determine the
properties of the plane wave as follows.  First, the propagation
direction $\hat{n}$ is the zero eigenvector of $h_{ij}(t,\vec{0})$.
(Actually, this only determines $\hat{n}$ up to a sign, since
$-\hat{n}$ is also a zero eigenvector \footnote{Any additional
  information, {\it e.g.} even crude directional information from
  triangulation, should break this $\pm\hat{n}$ degeneracy; I thank
  Kiyoshi Masui for this observation.}.)  Then, by projecting
$h_{ij}(t,\vec{0})$ onto the polarization tensors
$P_{A}^{ij}(\hat{n})$, we obtain the two polarization waveforms:
$h_{A}(t)= [P_{A}^{ij}(\hat{n})]^{\ast}h_{ij}(t,\vec{0})$.  We started
with $6$ functions of time $h_{ij}(t,\vec{0})$; so far we have used
these to completely determine the properties of the gravitational
plane wave [two functions $h_{A}(t)$ and two angles to specify
$\hat{n}$]; and the remaining information may now be used to perform
the following cross checks that we are really observing a standard
gravitational plane wave:
\begin{subequations}
  \begin{eqnarray}
    \label{angles}
    d\hat{n}/dt&=&0 \\
    \label{Det}
    {\rm Det}[h_{ij}(t,\vec{0})]&=&0 \\
    \label{Tr}
    {\rm Tr}[h_{ij}(t,\vec{0})]&=&0.
  \end{eqnarray}
\end{subequations}

Now let us turn to the noise $n_{\alpha}^{}(t)$, which we model as
stationary and gaussian, with zero mean, so that it is characterized
by its correlation function $C_{\alpha\beta}(T)$ or, equivalently, its
spectral density $S_{\alpha\beta}(f)=\tilde{C}_{\alpha\beta}(f)$:
\begin{subequations}
  \begin{eqnarray}
    C_{\alpha\beta}(T)&=&\overline{n_{\alpha}(t+T)n_{\beta}(t)} \\
    \delta(f-f')S_{\alpha\beta}(f)&=&\overline{\tilde{n}_{\alpha}^{\ast}(f)\tilde{n}_{\beta}(f')}.
 \end{eqnarray}
\end{subequations}
$S_{\alpha\beta}(f)$ induces a natural inner product on the space of
signals (or noise) in the network:
\begin{equation}
  (g^{(1)}|g^{(2)})=\int_{-\infty}^{+\infty}df\,
  \tilde{g}_{\alpha}^{(1)\ast}(f)\,
  [S^{-1}(f)]^{\alpha\beta}\,\tilde{g}_{\beta}^{(2)}(f).
\end{equation}
A particular noise fluctuation has probability proportional to ${\rm
  exp}[-(n|n)/2]$ and the expected signal-to-noise ratio (SNR) of a
gravitational wave is $(h|h)^{1/2}$.  If a gravitational wave signal
(which depends on various parameters $\theta_{k}$) is detected, and
the likelihood function may be approximated as a gaussian $\propto{\rm
  exp}[-(1/2)\theta_{k}\Gamma^{kl}\theta_{l}]$ near its peak, then the
expected inverse covariance matrix is
\begin{equation}
  \Gamma^{kl}=\left(\frac{\partial h}{\partial \theta_{k}}\Big|
    \frac{\partial h}{\partial\theta_{l}}\right).
\end{equation}

Consider a network of identical uncorrelated detectors:
$W_{\alpha}(T)=W(T)$ and $S_{\alpha\beta}(f)=
S(f)\delta_{\alpha\beta}$.  Such a network is a ``perfect porcupine''
if $(h|h)$ simplifies to the form
\begin{equation}
  (h|h)=C\int_{-\infty}^{+\infty}df\frac{|\tilde{W}(f)|^{2}}{S(f)}
  \left[\sum_{A=1}^{2}\left|\tilde{h}_{A}(f)\right|^{2}\right]
\end{equation}
where $C$ is a constant.  This says that the network's gravitational
wave sensitivity is independent of the direction or polarization of
the wave.  

Perfect porcupines also have other nice properties.  First, as we
shall see, they: (i) determine the propagation direction $\hat{n}$ and
both polarization waveforms $h_{A}(t)$ of a gravitational plane wave,
as described above; and (ii) permit the systematic checks
(\ref{angles}, \ref{Det}).  In addition, perfect porcupines built from
one-arm detectors (like AGIS \cite{Dimopoulos:2008sv}) {\it also}
permit the {\it other} systematic check (\ref{Tr});
we return to this point in the Discussion.  Next recall that
$h_{\alpha}(t)$ depends on two sorts of parameters: (i) the two angles
in $\hat{n}$, which we denote by the labels $\mu$ and $\nu$; and (ii)
all other parameters (such as the masses and spins and inclinations in
an inspiraling binary black hole) which we denote by the label
$\sigma$.  With this notation, if we choose angular coordinates that
are ``nice'' near the point $\hat{n}$, in the sense that they run
along two perpendicular great circles through $\hat{n}$ (like ordinary
polar coordinates $\theta$ and $\phi$ at the equator), then we find:
\begin{subequations}
  \begin{eqnarray}
    \Gamma^{\mu\nu}&=&(h|h)\delta^{\mu\nu}={\rm SNR}^{2}
    \delta^{\mu\nu} \\
    \Gamma^{\mu\sigma}&=&0.
  \end{eqnarray}
\end{subequations}
This says that the expected uncertainties in the two angular
coordinates of the source are equal to each other, mutually
uncorrelated, and independent of $\hat{n}$; and also that they are
uncorrelated with the uncertainties in all of the other parameters
characterizing the source.  Furthermore, the perfect porcupine's
angular resolution $\delta\theta=1/{\rm SNR}$ should be compared with
the angular resolution from triangulation:
$\delta\theta\sim(\lambda/L)(1/{\rm SNR})$, where $\lambda$ is the
gravitational wavelength and $L$ is the distance between the detectors
in the network.  Thus, in the (porcupine) regime where $L$ is short
relative to $\lambda$, the angular resolution of a perfect porcupine
is parametrically better than the angular resolution from
triangulation.

\section{Networks of one-arm detectors}

One-arm detectors (such as AGIS \cite{Dimopoulos:2008sv}) have antenna
projectors of the form:
\begin{equation}
  A_{ij}^{\alpha}=\hat{m}_{i}^{\alpha}\hat{m}_{j}^{\alpha}.
\end{equation}

The minimal perfect porcupine built from such detectors has 6 arms,
oriented along the 6 directions connecting opposite vertices of a
regular icosahedron (or, equivalently, the 6 directions connecting
opposite faces of a regular dodecahedron):
\begin{equation}
  \label{icosahedral6}
  \hat{m}^{\alpha}\!=\!\left\{
    \begin{array}{ll}
      \{0,0,1\} & (\alpha\!=\!0) \\
      \sqrt{4/5}\left\{{\rm cos}\frac{2\pi\alpha}{5},{\rm
          sin}\frac{2\pi\alpha}{5},\frac{1}{2}\right\} &
      (\alpha\!=\!1,\ldots,5)
    \end{array}
  \right.
\end{equation}
This network has $C=4/5$.  These 6 detectors are independent in the
sense of (\ref{grammian}); so in addition to determining the direction
of a gravitational plane wave $\hat{n}$, and both polarization
waveforms $h_{A}(t)$, they also permit the three cross checks
(\ref{angles}, \ref{Det}, \ref{Tr}) to be performed.

A larger perfect porcupine has 10 detectors, oriented along the 10
directions connecting opposite vertices of a regular dodecahedron (or,
equivalently, the 10 directions connecting opposite faces of an
icosahedron):
\begin{equation}
  \label{icosahedral10}
  \hat{m}^{\alpha}_{\pm}\!=\!\frac{2\varphi^{\pm1/2}}{\sqrt{3\sqrt{5}}}\left\{
    {\rm cos}\frac{2\pi\alpha}{5},{\rm sin}\frac{2\pi\alpha}{5},\frac{1\mp\varphi^{\mp1}}{2}\right\}
\end{equation}
where $\alpha=1,\ldots,5$, $\varphi=(1+\sqrt{5})/2$ is the golden
ratio, and $C=4/3$.  Of course, these 10 detectors are not independent
in the sense of Eq.~(\ref{grammian}): they measure all 6 independent
components of $h_{ij}$, but do so redundantly.  
All else being equal, this redundancy makes the 10-arm perfect
porcupine better (but also more expensive) than its 6-arm counterpart:
the redundancy allows us to cross correlate two noisy data streams
which contain the same gravitational wave signal, but different and
uncorrelated noise, and thereby extract the true signal better than we
could from either data stream individually.

An even larger and more redundant perfect porcupine consists of 15
detectors, oriented along the 15 directions connecting opposite edges
of a regular icosahedron (or, equivalently, the 15 directions
connecting opposite edges of a regular dodecahedron).  This has $C=2$.

If one relaxes the requirement that all of the detectors be identical,
more options become available.  For example, for $k\geq5$, the
configuration
\begin{equation}
  \hat{m}^{\alpha}\!=\!\left\{
    \begin{array}{ll}
      \{0,0,1\} & (\alpha\!=\!0) \\
      \sqrt{4/5}\left\{{\rm cos}\frac{2\pi\alpha}{k},{\rm
          sin}\frac{2\pi\alpha}{k},\frac{1}{2}\right\} &
      (\alpha\!=\!1,\ldots,k)
    \end{array}
  \right.
\end{equation}
with
\begin{equation}
  W_{\alpha}(T)\!=\!\left\{
    \begin{array}{ll}
      \sqrt{k/5}W(T) & (\alpha\!=\!0) \\
      W(T)& (\alpha\!=\!1,\ldots,k).
    \end{array}
  \right.
\end{equation}
is a perfect porcupine with $C=4k/25$.  The $k=5$ case is just the
minimal perfect porcupine (\ref{icosahedral6}).

\section{Networks of two-arm detectors}

Two-arm detectors (Michelson interferometers like LIGO/VIRGO, with
equal and orthogonal arms) have antenna projectors of the form
\begin{equation}
  P_{ij}^{\alpha}=\frac{1}{\sqrt{2}}(\hat{p}^{\alpha}_{i}\hat{p}^{\alpha}_{j}
  -\hat{q}^{\alpha}_{i}\hat{q}^{\alpha}_{j}).
\end{equation}

If we define $\kappa=\frac{1}{2}{\rm arccos}\sqrt{3/5}$ and
\begin{equation}
  \begin{array}{rcl}
    \hat{a}^{\alpha}&=&\left\{-{\rm sin}\frac{2\pi\alpha}{N},+{\rm
        cos}\frac{2\pi \alpha}{N},0\right\} \\
    \hat{b}^{\alpha}&=&\left\{-{\rm cos}\frac{2\pi\alpha}{N},-{\rm
        sin}\frac{2\pi\alpha}{N},\sqrt{2}\right\}\sqrt{1/3}
  \end{array}
\end{equation}
then, for $\alpha=1,\ldots,N$ and $N\geq5$, the configuration
\begin{equation}
  \begin{array}{rcl}
    \hat{p}^{\alpha}&=&+{\rm cos}\kappa\,\hat{a}^{\alpha}+{\rm
      sin}\kappa\,\hat{b}^{\alpha} \\
    \hat{q}^{\alpha}&=&-{\rm sin}\kappa\,\hat{a}^{\alpha}+{\rm
      cos}\kappa\,\hat{b}^{\alpha}
  \end{array}
\end{equation} 
is a perfect porcupine with $C=N/5$.  This network measures the 5
traceless components of $h_{ij}$ (with greater redundancy when $N$ is
larger).  It therefore determines the direction $\hat{n}$ of a
gravitational plane wave, and both polarization waveforms $h_{A}(t)$;
and it permits the first two cross checks (\ref{angles}, \ref{Det}),
but not the third (\ref{Tr}).  When $N=5$ this is the minimal perfect
porcupine.

Another perfect porcupine is worth mentioning.  Consider the 15
directions connecting opposite edges of a regular icosahedron (or,
equivalently, the 15 directions connecting opposite edges of a regular
dodecahedron).  These 15 directions separate into 5 orthonormal
triads.  From each orthonormal triad, we can select 3 different
orthonormal pairs $\{\hat{p}_{\alpha},\hat{q}_{\alpha}\}$.  In this
way, we obtain a perfect porcupine with 15 detectors, and $C=3$.

\section{Discussion}

We have seen that a perfect porcupine built from one-arm detectors
will monitor ${\rm Tr}[h_{ij}]$.  As mentioned above, the vanishing of
this channel is a check that one is observing standard gravitational
waves, as opposed to noise, or something more exotic.  Alternatives to
general relativity often give the graviton a zero-helicity component;
and zero-helicity gravitational waves would show up as fluctuations in
${\rm Tr}[h_{ij}]$.  Also, as one goes underground, and to lower
frequencies, the coherence length of the fluctuations in the Newtonian
gravitational potential becomes longer: if the various detectors in a
perfect porcupine can be placed sufficiently close together, so that
they all see the same (or similar) fluctuations in the Newtonian
potential, then these fluctuations will also look like fluctuations in
${\rm Tr}[h_{ij}]$, and monitoring this channel may even be helpful in
subtracting gravity gradient noise.  On the other hand, if it is
necessary, {\it e.g.}, to build the detectors parallel to Earth's
local gravitational field (so that the various detectors in the
porcupine must be located at widely separated points on the Earth),
then this method for removing gravity gradient noise won't work.

I have focused on a perfect porcupine's ability to measure a single
plane wave; but this analysis also applies to a sum of many plane
waves, as long as they are separable in the time-frequency (or
template) domain.  For example, two plane waves of different
frequencies are not a problem (they may be cleanly separated in
frequency space, and then handled independently); but two plane waves
moving in different directions at the {\it same} frequency would be a
problem: the cross check (\ref{Det}) and the porcupine's
direction-finding algorithm would fail.  Fortunately, although source
``blending'' (non-separability) does occur for LISA sources when
$f\lesssim10^{-3}~{\rm Hz}$, it is unlikely that a ground-based
perfect porcupine would reach sufficiently low frequencies and
sufficiently good sensitivities for this to be a practical concern.

An earth-bound porcupine must contend with (in order of importance)
the spin of the Earth, the Moon's orbit around the Earth, and the
Earth's orbit around the Sun; but above we considered perfect
porcupines which move in straight line, without rotation.  As long as
we are looking at gravitational waves with periods much shorter than a
day (which is the realistic case), then this is a good starting point,
for the same reason that for many purposes LIGO/VIRGO may be modeled
as moving through flat space: over timescales containing many wave
cycles, the detectors {\it are} moving along a nearly straight
non-rotating trajectory.  Nevertheless, it is important to extend the
above analysis to include the rotation of the Earth, etc., especially
for the purposes of studying sources that are detectable over
timescales longer than a day.  It will also probably be best to align
one of the perfect porcupine's symmetry axes with the Earth's rotation
axis, if possible.

Finally, the considerations in this paper were motivated by recent
developments (outlined in the Introduction), but I should also mention
a more futuristic possibility: we may eventually be led to build
low-frequency gravitational wave detectors on the Moon, where the
seismic and gravity gradient noise levels are much lower than on
Earth.  This would again lead us to porcupines.

I acknowledge support from the CIFAR JFA.

\end{document}